Tidal Heating of Terrestrial Extra-Solar Planets

and Implications for their Habitability

by

Brian Jackson, Rory Barnes, & Richard Greenberg

Abstract

The tidal heating of hypothetical rocky (or terrestrial) extra-solar planets spans a wide range of values depending on stellar masses and initial orbits. Tidal heating may be sufficiently large (in many cases, in excess of radiogenic heating) and long-lived to drive plate tectonics, similar to the Earth's, which may enhance the planet's habitability. In other cases, excessive tidal heating may result in Io-like planets with violent volcanism, probably rendering them unsuitable for life. On water-rich planets, tidal heating may generate sub-surface oceans analogous to Europa's with similar prospects for habitability. Tidal heating may enhance the outgassing of volatiles, contributing to the formation and replenishment of a planet's atmosphere. To address these issues, we model the tidal heating and evolution of hypothetical extra-solar terrestrial planets. The results presented here constrain the orbital and physical properties required for planets to be habitable.

I. Introduction

As observational technology and techniques improve, it is anticipated that Earth-like planets may be found orbiting stars other than the Sun. Such a discovery would be a milestone in the search for extra-terrestrial life. One technique for finding extra-solar planets, the radial-velocity observations of distant stars, is sensitive to Earth-mass planets if they orbit close to inactive, low-mass stars. Also, since M stars are less luminous than more massive stars, the conventional habitable zone (HZ, as defined by Selsis *et al.* 2007; see also Kasting *et al.* 1993) lies closer to the star than for a star of a greater mass. Planets must be closer to their M star host in order to have a surface temperature suitable for life. Taken together, these criteria suggest that the first Earth-like, habitable planets may be discovered around M stars (Tarter *et al.* 2007).

The effects of tides on rocky (or "terrestrial") planets close to their host stars are likely to be important (Mardling & Lin 2004; Mardling 2007; Barnes *et al.* 2008; Jackson *et al.* 2008b). If such a planet is on an eccentric orbit, the dissipation of tidal energy within the planet will tend to circularize the planet's orbit and probably lead to substantial internal heating. As we show in this paper, such heating can have important consequences for the habitability of these bodies.

Jackson *et al.* (2008b) considered the effects of tidal heating on observed extra-solar planets. The tidal heating that accompanies orbital evolution was considered for jovian-scale extra-solar planets, where it may have major effects on physical properties. For example, the radii of some transiting extra-solar planets are observed to be larger than theoretical modeling predicts (*e.g.*, Bodenheimer *et al.* 2003; Burrows *et al.* 2007), but tidal heating of the interiors of these planets may have been enough to pump up their radii (Jackson *et al.* 2008b). Jackson et al. also showed that tidal heating of known extra-solar planets, if they were terrestrial, may be enough to completely melt some planets. Thus many extra-solar planets could be at least as volcanically active as Jupiter's moon Io, the most volcanically active body in our solar system, with massive volcanic plumes and frequent global resurfacing (Peale *et al.* 1979; McEwen *et al.* 2004). In fact, tidal heating of the innermost planet in the recently discovered system HD 40307 (Mayor et al. 2008) may exceed Io's, suggesting that, if the planet is rocky, it may be volcanically active (Barnes et al. 2008b).

On the Earth, internal heating drives convection in the mantle, and thus, contributes to the process of plate tectonics, although this process is not completely understood (O'Neill & Lenardic 2007). For the Earth, tidal heating is negligible, and adequate heating is provided by decay of radionuclides. Plate tectonics helps stabilize the planet's atmosphere and surface temperature over hundreds of millions of years (Walker *et al.* 1981). Because a stable surface temperature is probably a prerequisite for life, plate tectonics may be required for a planet to be habitable (Ragenauer-Lieb *et al.* 2001).

These considerations suggest that, in order to be habitable, terrestrial planets require a significant source of internal heating. However, models of *in situ* planet formation predict that terrestrial planets that form around M stars would have masses much less than the Earth (Raymond *et al.* 2007). On the other hand, shepherding of terrestrial planets by migrating gas giants may result in larger terrestrial planets (Raymond *et al.* 2008). Thus, if *in situ* formation of terrestrial planets is common, many terrestrial planets we find may have too little radiogenic heating to drive long-lived plate tectonics (Williams *et al.* 1997). However, even without radiogenic heating, as we show, tides may provide adequate internal heating and hence may be critical in determining planetary habitability.

In many cases tidal heating may even be too great to allow habitability. For example, tides on planets with masses comparable to or larger than the Earth's may generate more internal heat (per unit surface area) than Io, even for small eccentricities. Even Planets with masses smaller than Mars may experience similar heating with eccentricities ~ 0.3, the average value for observed extra-solar planets. Thus, many extra-solar terrestrial plants may be extremely volcanic, perhaps enough to preclude habitability.

Tidal heating may also be critical for creation and maintenance of the atmospheres of terrestrial planets. As heat induces internal convection within the mantles of these planets, volatiles trapped in the mantle may be outgassed, feeding the atmosphere. However, in the HZ of M stars, atmospheres may be depleted by vigorous stellar activity (West et al. 2008) or impact erosion (Melosh & Vickery 1989). Tides may be crucial to drive adequate outgassing to replenish the atmosphere, enabling the planet to remain habitable.

In other cases, planets may be habitable even without an atmosphere. Tidal heating of an icy planet may generate a habitable subsurface ocean, analogous to Jupiter's moon Europa. For these planets, an atmosphere may not be necessary for habitability, if, as proposed for Europa, life could exist below the icy surface.

To determine the impact of tidal heating on the geophysics, surfaces and atmospheres of possible extra-solar terrestrial planets, and thus ultimately on their habitability, we model the tidal evolution of a suite of hypothetical terrestrial planets in the HZ of their host stars. We explore a range of planetary and stellar masses and orbital parameters, using a classical model for tidal evolution (Goldreich & Soter 1966; Kaula 1968; Jackson *et al.* 2008a) and track the tidal evolution for each planet over billions of years. Tides raised on the planet tend to damp down the orbital eccentricity *e* at the same time as they heat the planet and reduce the semi-major axis *a*. If a tidally evolving terrestrial planet begins with a non-zero eccentricity, it may take billions of years for tides to damp that eccentricity to near-zero, even without gravitational interactions with other planets. Consequently, we find tidal heating can be very long-lived, even for an isolated planet.

A common approach in the literature has been to consider a so-called "timescale" for damping as a way to estimate the effects of tides (*e.g.*, Rasio *et al.* 1996). However, that approach can be inaccurate because the tidal evolution of *a* and *e* are strongly coupled. Changes are much more complex than simple exponential decay, so characterizing the evolution with a timescale is inappropriate (Jackson *et al.* 2008a). Eccentricity values can remain large for longer than previously recognized, which can extend the duration of tidal heating. This result also means that observers looking for extra-solar planets close to their host stars should not discount *a priori* indications of a non-zero *e*, even if there are no other planets in a system.

In the next section, we give the details of our model. In Section III, we present the results of our integrations, while in Section IV, we discuss the implications of these results. In Section V, we summarize our results.

II. Methods

To calculate the tidal heating of hypothetical extra-solar terrestrial planets, we performed a suite of integrations of the tidal evolution equations, over a range of stellar and planetary masses and a range of orbital elements. We considered stellar masses $M_s$ from 0.1 to 0.5 solar masses ($M_{sun}$), which is the range of masses for M stars. We sampled planetary masses $M_p$ from 0.01 to 10 Earth masses ($M_{Earth}$). For each star, we assumed a radius based on the star's mass, according to Gorda & Svechnikov (1998). For each planet, we assumed a radius ($R_p$) based on geophysical models by Sotin *et al.*

(2007): $R_p = M_p^{0.27}$ for $1 \leq M_p \leq 10$ and $R_p = M_p^{0.3}$ for $0.1 \leq M_p < 1$, where $R_p$ is in Earth radii ($R_{Earth}$).

We also considered a range of initial orbital eccentricities $e_0$ from 0.01 to near unity, a range similar to known extra-solar planets (*e.g.*, Butler *et al.* 2006). Each planet was also assigned an initial orbital semi-major axis $a_0$ in the middle of its host star's HZ, as defined by Barnes *et al.* (2008) who modified the definition from Selsis *et al.* (2007). Selsis *et al.* related a star's HZ to its mass, a planet's cloud cover, and a planet's semi-major axis. Barnes *et al.* (2008) expanded the relationship to include the effects of non-zero eccentricities on the range of habitable semi-major axes, using the results from Williams & Pollard (2002). For each planet, we chose the HZ range assuming no cloud cover (Selsis *et al.* 2007). Fig. 1 shows the dependence of $a_0$ (at the middle of the HZ) on $e_0$ from Barnes *et al.* (2008, Eq. 8 & 9). For large $e_0$, the HZ is shifted outward from any given star, so we choose a larger $a_0$.

Each planet was assumed to have zero obliquity and a rotation that had spun-down so early that exchange of angular momentum between the planet's orbit and its rotation is not a factor. Such spin-down probably takes < 1 Myr in most cases (Peale 1977). For planets with non-zero eccentricity, this assumption means that the planet's rotation period may be shorter than the orbital period (Greenberg & Weidenschilling 1984; Barnes *et al.* 2008).

The tidal evolution formulae adopted here were derived under the assumption that terms of higher order than $e^2$ are negligible (Jeffreys 1961; Kaula 1968; Goldreich & Soter 1966). Higher order terms may be important and might change the rates of tidal heating and evolution. Tidal models that incorporate higher order corrections exist, but involve numerous assumptions about dissipation processes within planetary bodies, which are complicated and uncertain (*e.g.,* Mardling & Lin 2002; Efroimsky & Lainey 2007; Ogilvie & Lin 2004). Future work may provide more accurate tidal models, but, for now, we used the classical formulation, as compiled by Jackson *et al.* (2008a) from Goldreich & Soter (1966) and Kaula (1968):

$$\frac{1}{e}\frac{de}{dt} = -\left(\frac{63}{4}(GM_s^3)^{1/2}\frac{R_p^5}{Q_p'M_p} + \frac{171}{16}(G/M_s)^{1/2}\frac{R_s^5 M_p}{Q_s'}\right)a^{-13/2} \quad (1)$$

$$\frac{1}{a}\frac{da}{dt} = -\left(\frac{63}{2}(GM_s^3)^{1/2}\frac{R_p^5}{Q_p'M_p}e^2 + \frac{9}{2}(G/M_s)^{1/2}\frac{R_s^5 M_p}{Q_s'}\right)a^{-13/2} \quad (2)$$

where $G$ is the gravitational constant, $Q_s'$ is the stellar dissipation parameter, $Q_p'$ is the planetary dissipation parameter. For the stars we assume $Q_s' = 10^{5.5}$ (Jackson *et al.* 2008a). For terrestrial planets, we assume $Q_p' = 500$, which incorporates a conventional value for the planetary Love number of 0.3 (Dickey *et al.* 1994; Mardling & Lin 2004), although actual values could vary, depending on structure and composition.

The tidal heating rate of the planet is equal to the portion of the loss of orbital energy due to tides raised on the planet, which is proportional to the first term in the equation for *da/dt* (Eq. 1). Thus the heating rate is given by:

$$h = \left(\frac{63}{16\pi}\right)\frac{(GM_s)^{3/2} M_s R_p^3}{Q_p'}a^{-15/2}e^2. \quad (3)$$

Here $h$ is expressed in terms of the internal heating per unit surface area of the planet for convenience in consideration of various effects on planetary surfaces later in this paper.

We numerically integrated the tidal equations, Eqs. (1) and (2), forward in time, beginning at $a_0$ and $e_0$, and calculated the heating that accompanies the evolution, using

Eq. (3). For the range of parameters considered here, we find that the effects of the tides raised on the star by the planet (given by the second terms in Eqs. (1) and (2)) are small. Because tides on the star don't contribute, the orbital angular momentum is approximately conserved.

As tides cause a planet's orbit to evolve, $h$ often passes through a maximum. Initially tides reduce $a$ without significantly reducing $e$, and $h$ increases. As $a$ becomes smaller, the rate of tidal circularization increases, and eventually $e$ decreases enough that $h$ decreases as well.

The maximum value of $h$ occurs where $dh/dt = 0$. Differentiating Eq. (3) and setting $dh/dt$ equal to zero yields:

$$2e\frac{de}{dt}a^{-15/2} - \frac{15}{2}e^2 a^{-17/2}\frac{da}{dt} = 0 \qquad (4)$$

With conservation of orbital angular momentum $l = (G\,M_s\,a\,(1-e^2))^{1/2}$, we have

$$\frac{d(l^2)}{dt} = \frac{da}{dt}(1-e^2) - 2ae\frac{de}{dt} = 0 \qquad (5)$$

which yields

$$\frac{da}{dt} = \frac{2ae}{1-e^2}\frac{de}{dt}. \qquad (6)$$

Substituting $da/dt$ from (6) into (4) yields $e = \sqrt{2/17} \approx 0.34$. Thus whenever $h$ reaches a maximum, $e$ is equal to this particular numerical value.

III. Results

Figs. 2 and 3 show the tidal heating rates and tidal evolution of $e$ and $a$ over time for planets of various masses and various values of $e_0$ in the HZ of an M star with a mass of 0.1 $M_{Sun}$ and 0.2 $M_{Sun}$, respectively. In none of our cases do we find that the tidal evolution carries the planet out of the HZ. (See Barnes et al. 2008a for a discussion of this possibility.)

In Fig. 2, qualitative, as well as quantitative, differences appear in the history of $h$, depending on the value of $e_0$. Because $h$ peaks when $e \approx 0.34$, planets that begin with $e_0 < 0.34$ have their heating rates drop monotonically with time. For larger $e_0$ (above 0.34), the initial and peak values of $h$ are larger. However, for $e_0 > 0.6$, we see that the initial value of $h$ is smaller, and, while the peak in $h$ is larger, it occurs later. It may seem surprising that these largest values of $e_0$ give less initial heating. However, this effect arises from our requirement (for this study) that $a_0$ be in the middle of the HZ. From Fig. 1, we see that $a_0$ is considerably larger for larger $e_0$. The larger $a_0$ means the initial value of $h$ is reduced (Eq. 3). Similarly the initial value of $de/dt$ is smaller (Eq. 1), so it takes longer for $e$ to reach 0.34, the value at which $h$ reaches a maximum.

Fig. 2 also shows how $h$ depends on $M_p$. Since $h$ is proportional to $R_p^3$, it depends roughly linearly on $M_p$ (Sotin et al. 2007). For example, for $M_p = 0.3$ $M_{Earth}$ and $e_0 = 0.1$, (Fig. 2a), the initial value of $h$ is about 0.3 W/m$^2$. By comparison, for a planet three times as massive (Fig. 2b), $h$ begins at about 1 W/m$^2$. For the range of $M_p$ we consider here, the orbits of larger planets evolve more quickly than those of smaller planets, and so are eccentric for less time.

In Fig. 3, we show the tidal heating for a planet with $M_p = 10$ $M_{Earth}$ orbiting a star with $M_s = 0.2$ $M_{Sun}$. Since stars with larger masses are more luminous, planets in the HZ of more massive stars are farther from the star. The rate of tidal evolution for planets in the HZ of these stars is much smaller since $a$ is larger. Thus, while the heating rates for these planets may be significant, they change very slowly. For smaller $M_p$, $h$ remains essentially unchanged from its initial value, which depends on $M_p$, $M_s$ and the initial eccentricity.

The contours in Figs. 4 and 5 show how $h$ depends on $M_p$, $M_s$ and $e$, at a fixed moment in time for a planet in the center of the HZ. (The nearly constant heating rates for planets with $M_p < 10$ $M_{Earth}$ orbiting a star with $M_s \geq 0.2$ $M_{Sun}$ can be read off Figs. 4 and 5.) In both figures, $h$ increases as $e$ increases from zero, but then decreases as $e$ becomes larger than ~0.6. The increase with $e$ is due to the $e$-dependence in Eq. 3, but the decrease in $h$ as $e$ gets very large arises, again, because the HZ moves farther from the star (see Fig. 1).

IV. Discussion

Tidal heating may affect a planet's habitability in various ways: (a) It may be great enough to drive plate tectonics for a length of time that depends sensitively on the host star and planet's masses and the planet's initial orbit. (b) It may be so great as to make the planet uninhabitably volcanic. (c) It may drive outgassing from the planet's interior, continually replenishing the planet's atmosphere against loss. (d) It may also be sufficient to produce a habitable subsurface ocean on an icy planet, or to mitigate life-challenging "snow-ball" conditions on a terrestrial planet.

Interpretation of the heating rates computed here is facilitated by comparison with known planetary bodies. The Earth emits ~ 0.08 W/m$^2$ of internal (radiogenic and primordial) heat (Davies 1999), which is apparently adequate for plate tectonics. For an estimate of the minimum heating required for tectonic activity, consider the geologic history of Mars. When Mars was last tectonically active, its radiogenic heat flux was ~ 0.04 W/m$^2$ (Williams *et al.* 1997), perhaps a minimum amount required for tectonics in a rocky planet. At the opposite extreme, Jupiter's moon Io emits 2 – 3 W/m$^2$ of internally generated heat from tides (Peale *et al.* 1979; McEwen *et al.* 2004), which results in extreme planet-wide volcanism, including active volcanic plumes and rapid resurfacing. Jupiter's moon Europa, a rocky body covered by 150 km of $H_2O$, may generate as much as 0.19 W/m$^2$, scaling from tidal heating on Io (O'Brien *et al.* 2002). Tidal heating in Europa maintains a sub-surface water ocean, and with that value of $h$, the surface ice would be only a few km thick (Greenberg 2005).

By comparing our calculations of $h$ for hypothetical planets to the above heating rates for known planets, we can estimate the range of masses and orbital elements that correspond to each of these physical effects. Of course, the geophysical processes discussed above are complicated, and the various processes may be very different on extra-solar planets than on planets in our solar system. For example, O'Neill & Lenardic (2008) show that plate tectonics may be less likely on terrestrial planets more massive than the Earth, even though they would probably experience more radiogenic heating. To conclude that these processes actually occur on a given planet requires more complete modeling than presented here. However, our calculations provide a starting point for such work by identifying ranges of possible planets for which tidal heating may be critical in establishing habitable conditions.

With reference to Figs. 4 and 5, we next consider how various heating-rate regimes in the ($e$, $M_p$, $M_s$) space affect prospects for habitability.

IV. a. Planets with Sufficient (but not Excessive) Tidal Heating for Plate Tectonics

First, we consider requirements for enough heating for plate tectonics, but not so much that there is devastating volcanism. As described above, this range is perhaps between ~0.04 and ~2 W/m$^2$. The light gray shading in Figures 4 and 5 indicates the range of parameters for which $h$ lies in this range. Figure 4 (a) and (b) show that tidal heating might drive plate tectonics on even a Mars-sized planet ($M_p = 0.3$ M$_{Earth}$) orbiting a star of mass 0.1 M$_{Sun}$ even with $e$ as small as 0.04. For planets around larger stars (Figure 4c), larger eccentricities are required for sufficient internal heating. Figure 5 shows that sufficient but not excessive heating also requires a specific range of $M_s$ and $e$.

In order to consider whether life might develop and evolve on a given planet, it is also useful to estimate how long $h$ lies in the acceptable range of values as the planet tidally evolves. Figures 6 and 7 show how long $h$ lies in this range between too little for plate tectonics and so much that volcanism is intolerable. It is important to note that, for many of our modeled planets, the initial value of $h$ is above the acceptable range, and only later, when the eccentricity is sufficiently small, does $h$ enter the acceptable range. For these cases, the time during which $h$ lies in the acceptable range, as depicted in Figs. 6 and 7, begins several Gyrs after the start of the simulation.

Fig. 6 (a) shows that with $M_s = 0.1$ M$_{Sun}$ and $M_p \leq 8$ M$_{Earth}$, increasing $e_0$ results in increasing duration of acceptable $h$ values until $e_0$ reaches a critical value between 0.3 and 0.6 (depending on $M_p$), at which point the duration drops again. For these largest values of $e_0$, the orbital circularization is fast. On the other hand, for $M_s = 0.2$ M$_{Sun}$ (Fig. 6b), the HZ is far enough from the star that planets can remain in the acceptable h range longer than 10 Gyr, even in excess of 100 Gyr.

Figure 7 shows how the heating lifetime depends on $M_s$. There is an abrupt transition from planets that never undergo habitable heating to planets with heating lifetimes greater than 10 Gyr, again as a result of the dependence of $a_0$ on $M_s$. Moving from panel (a) to (d), $M_p$ increases from 0.3 to 10 $M_{Earth}$. An island of small heating lifetimes centered around $e_0 = 0.7$ and small $M_s$ grows larger with larger $M_p$, as the initial heating rate is larger. After billions of years, however, the eccentricities of these large planets drop, and the heating rates eventually pass through the habitable range.

Figures 6 and 7 suggest that the habitability of a terrestrial planet will require tight constraints on the planet's orbital and physical parameters to keep $h$ in the acceptable range for a reasonable amount of time. These calculations suggest that for plausible initial conditions, even small terrestrial planets would remain habitable for longer than the age of the Earth.

IV. b. Planets with Excessive Volcanism

For $h \geq 2$ W/m$^2$, by analogy with Io, habitability may be unlikely, because tidal heating may drive volcanic and tectonic activity too vigorous to allow life. The dark gray shading on Figures 4 and 5 shows the range of $M_p$, $M_s$ and $e$ for which $h$ exceeds 2 W/m$^2$. For $M_s = 0.1$ $M_{Sun}$ (Fig. 4 (a) and (b)), a very wide range of $e$ and $M_p$ result in this overheating. However, for $M_s = 0.2$ $M_{Sun}$ (Fig. 4 (c)), planets in the HZ are unlikely to experience heating as large as Io's.

Fig. 5 gives another perspective on the range of $M_p$, $M_s$ and $e$ values that give overheating. Planets with $M_p = 0.3$ $M_{Earth}$, for example, only experience heating comparable to Io's when orbiting a very small star, *e.g.* with $M_s \sim 0.1$ $M_{Sun}$. For a larger planet in the HZ of such a small star, $h$ could be far greater, *e.g.* with $M_p = 10$ $M_{Earth}$, $h > 30$ W/m$^2$ (Fig. 5d).

Given the wide range of masses and eccentricities that potentially give rise to extreme volcanism, we might expect that many terrestrial planets will be too volcanically active for life. In fact, as Fig. 5 (d) shows, the most massive terrestrial planets may also be the most heated and thus the most volcanically active. Since the first extra-solar terrestrial planet that is likely to be confirmed will probably be much more massive than the Earth, we might expect it will be volcanically active. Such volcanic activity may be recognizable in the planet's atmospheric transmission spectrum, similar to Io, whose tenuous atmosphere is largely made of sulfur (Geissler *et al.* 1999).

Our calculations also suggest the interesting possibility that a tidally evolving planet may pass through alternate epochs of sufficient heating and excessive heating, even without interactions with other planets. Consider the $e_0 = 0.9$ curve in Figure 2 (a). The initial value of $h$ is ~ 1 W/m$^2$, sufficient for plate tectonics but not excessive. As time progresses and tides bring the planet closer to the star (smaller $a$), $h$ slowly grows larger than 2 W/m$^2$. Eventually after tens of billions of years (beyond the time shown in Fig. 2a), the orbit will circularize so that $h$ drops below 2 W/m$^2$ again. Thus, such a planet may begin habitable, then become too violently volcanic, and then, tens of billions of years later, become habitable again. Perhaps life on such a planet would arise in two separate epochs, billions of years apart.

IV. c. Tidal Heating and Planetary Atmospheres

Tidal heating may also help some extra-solar terrestrial planets to retain atmospheres long enough to allow life to develop and evolve. In some cases such help may be essential, because at least two other processes might otherwise quickly remove the atmospheres. Planets with masses comparable to Mars' will have such low surface gravity that impactors may quickly erode their atmospheres. Melosh & Vickery (1989) found that impact erosion may have reduced the atmosphere of Mars to its current surface pressure from an initial pressure of 1 bar over solar system history. The atmospheres of extra-solar terrestrial planets may also be depleted by stellar activity. Some M stars are so active (Hawley *et al.* 1996; Tarter *et al.* 2007; West et al. 2008) that coronal mass ejections (CMEs) could remove atmospheres from planets in the HZ. Lammer *et al.* (2007) showed that, without sufficient shielding by a planetary magnetosphere, an Earth-like planet with $a \leq 0.2$ AU may lose its entire atmosphere in a billion years, limiting prospects for long-term habitability and evolution. On the other hand, tidal heating of the planet may enhance internal convection and help to generate a magnetic field, which could shield the planet's atmosphere.

However, tidal heating may supply or replenish atmospheres that might otherwise be lost. For example, the radiogenic heating of the Earth is sufficient to induce outgassing of volatiles from the Earth's mantle (Williams *et al.* 1997). Papuc & Davies (2008) modeled the thermal and tectonic evolution of hypothetical terrestrial planets, and found that, for a wide range of planetary masses, volcanic outgassing may provide an atmosphere as massive as the Earth's in a few Gyrs.

However, those calculations considered radiogenic heating as the only heat source. Our calculations show that tidal heating may easily exceed radiogenic heating, so it may significantly increase the mantle outgassing rate, in many cases enough to resupply atmospheres that otherwise would be lost to erosion by impactors or CMEs. The habitability of terrestrial planets requires a suitable balance between internal and atmospheric processes, and tidal heating is potentially great enough that it may play a critical role.

IV. d. Ice-covered Planets

Tidal heating could also enhance the habitability of planets with large water components, by maintaining a liquid ocean just below the surface (Vance *et al.* 2007), even if the exposed surface is frozen. Such a planet would be similar to Europa, where (as described above) tides keep most of the water liquid under a thin ice shell. The sub-surface ocean might be suitable for life (Reynolds *et al.* 1983; Chyba 2000; Greenberg 2005). Europa has a mass of 0.008 $M_{Earth}$ and is composed of a rocky interior below a ~150-km-thick layer of water, mostly liquid due to tidal heating (Greenberg 2005). The chemicals required for life, including oxidants, may be generated on Europa's surface by energetic charged particles trapped in Jupiter's magnetosphere (Johnson *et al.* 2004). These substances may be transported into the sub-surface ocean by various processes (Greenberg 2005), including impacts (Chyba & Phillips 2002) or local melt-through of the surface ice (O'Brien *et al.* 2002). It has also been speculated that the tidal heating may also give rise to underwater volcanic vents that could support life directly with thermal energy, by analogy to deep ocean life on the Earth (Baross *et al.* 1982).

Tidal heating may be adequate to drive processes on similar extra-solar ice-covered worlds. Fig. 4 shows that planets with $M_p \sim 0.01$ $M_{Earth}$ can experience $h > 0.08$ W/m$^2$ with $e \sim 0.3$, and hence might support a subsurface ocean if there were water present. Also similar to Europa, energetic charged particles, in this case from stellar CMEs rather than from Jupiter's magnetosphere, might give rise to oxidants and other biologically useful substances. In this way, as for Europa, the energetic particles that endanger surface life on Earth-like worlds may enable sub-surface life under an icy crust. Whether a planet as volatile-rich as Europa could actually form around an M star is speculative (Lissauer 2007; Raymond *et al.* 2007), but if they do form, the same processes that may make Europa habitable may also make these planets habitable.

V. Conclusions

From our calculations, we can draw several important conclusions. Many extra-solar terrestrial planets may experience sufficient internal tidal heating to drive plate tectonics. Planets as small as Mars could maintain plate tectonics, even if radiogenic heating alone would be insufficient. Over a wide range of planetary and stellar masses and orbital elements, tidal heating may maintain plate tectonics for billions of years, as a planets orbit evolves and its eccentricity damps down. To maintain such long-lived geophysical activity, it is not necessary to invoke interactions with other planets to pump the eccentricity.

Under a different range of circumstances, tidal heating may be so great as to drive extreme, life-challenging volcanism. For example, we have shown how planets with masses comparable to or larger than the Earth may experience more tidal heating per unit surface area than Io. Because extreme heating may be common among close-in planets with masses larger than the Earth's, the first extra-solar terrestrial planets to be confirmed may prove to be inhospitable to life. The recently discovered planetary system around HD 40307 (Mayor et al. 2008) may represent the first test case for this hypothesis.

Tidal heating may enhance the prospects for life on other planets in various ways. It may enhance the outgassing of volatiles, counteracting the removal of gas by impact erosion and stellar CMEs. In that way tidal heat may be critical to maintaining atmospheres and thus habitability. With or without an atmosphere, if a planet is ice-covered, like Europa, tidal heating may give rise to a habitable subsurface ocean, protected from organism-damaging CMEs by the surface ice. At the same time, CMEs may actually enable some planets to be habitable by creating essential substances, such as oxidants.

Even with the many simplifying assumptions employed here, these results suggest a wide range of geophysical scenarios. As advancement is made in the understanding of the processes of tidal evolution, in modeling of the geophysics of hypothetical planets, and eventually in the discovery and characterization of actual terrestrial-type planets, these calculations will need to be revisited. In any case, the calculations here show that tidal heating has the potential to be a major factor in governing the internal structures, surfaces and atmospheres of extra-solar terrestrial planets. Accordingly, the effects of tidal heating must be given consideration when evaluating the habitability of such planets.

Acknowledgments

The authors would like to thank Lisa Kaltenegger, Sean Raymond and Christophe Sotin for useful discussions. This study was funded by NASA's Planetary Geology and Geophysics Program.

Figure Captions

Figure 1: For circular orbits ($e_0 = 0$), the orbital semi-major axis at the center of the HZ ($a_0$) depends only on the stellar mass $M_s$. For $e_0 > 0$, the value of $a_0$ shifts to larger values as shown.

Figure 2: Tidal heating histories and tidal evolution of $e$ and $a$ for planets of various masses with various values of initial eccentricity $e_0$ orbiting a star with mass 0.1 $M_{Sun}$. Parts (a) - (d) illustrate histories for planets with masses 0.3, 1, 3, and 10 $M_{Earth}$, respectively. These planets are assumed to be in the HZ of a star of mass 0.1 $M_{Sun}$. For all panels for a given $M_p$, solid lines correspond to $e_0 = 0.05$, dotted lines to 0.1, dashed to 0.3, dashed-dotted to 0.6, and dashed-dot-dot-dot to 0.9. The tidal heating rate is expressed in terms of $h$, the heating divided by the planet's surface area. For comparison, $h$ for Io and the Earth are 2 and 0.08 W/m$^2$, respectively.

Figure 3: Tidal heating histories and tidal evolution of *e* and *a* for a planet with mass $M_p$ = 10 $M_{Earth}$ with various values of $e_0$ orbiting a star with mass 0.2 $M_{Sun}$. Different linestyles correspond to the same values of $e_0$ as in Fig. 2. For smaller planetary masses, tidal heating is constant over 15 Gyr.

Figure 4: Contours of heating in W/m$^2$ for a range of $M_p$ and *e*. For panels (a) and (b), $M_s$ = 0.1 $M_{Sun}$, and $M_s$ = 0.2 $M_{Sun}$ for panel (c). Panel (b) shows a close-up of the bottom left corner of panel (a). White areas indicate $h < 0.04$ W/m$^2$. Light gray areas indicate 0.04 W/m$^2 \leq h \leq 2$ W/m$^2$ (adequate for plate tectonics) while dark gray indicates $h > 2$ W/m$^2$ (more than Io).

Figure 5: Contours of heating in W/m$^2$ for a range of $M_s$ and *e*. For panels (a) – (d), $M_p$ = 0.3, 1, 3, and 10 $M_{Earth}$. White and gray areas have the same meanings as in Fig. 4.

Figure 6: Duration of tidal heating in the "habitable range" (*i.e.* enough for plate tectonics, but not enough for sterilizing, Io-like volcanism) for various $M_p$ and *e*. For panels (a) and (b), the corresponding values of $M_s$ are 0.1 and 0.2 $M_{Sun}$. The contours of heating lifetimes are labeled in Gyrs, and contours levels for 1, 3, 5, 8, and 10 Gyrs are shown. In panel (b), since the change in tidal heating rate is very slow, planets that begin with sufficient tidal heating have sufficient heating for longer than 10 Gyr, longer than 100 Gyr in some cases. In the region shaded gray and labeled "never", planets start out with too little heating and always have too little heating.

Figure 7: Tidal heating lifetimes for various $M_s$ and $e$. Panels (a) – (d) correspond to $M_p$ = 0.3, 1, 3, and 10 $M_{Earth}$, respectively. Again, the contours of heating lifetimes are labeled in Gyrs and are shown for 1, 3, 5, 8, and 10 Gyr, and planets in the shaded gray region, labeled "never", never have high enough heating rates for habitability.

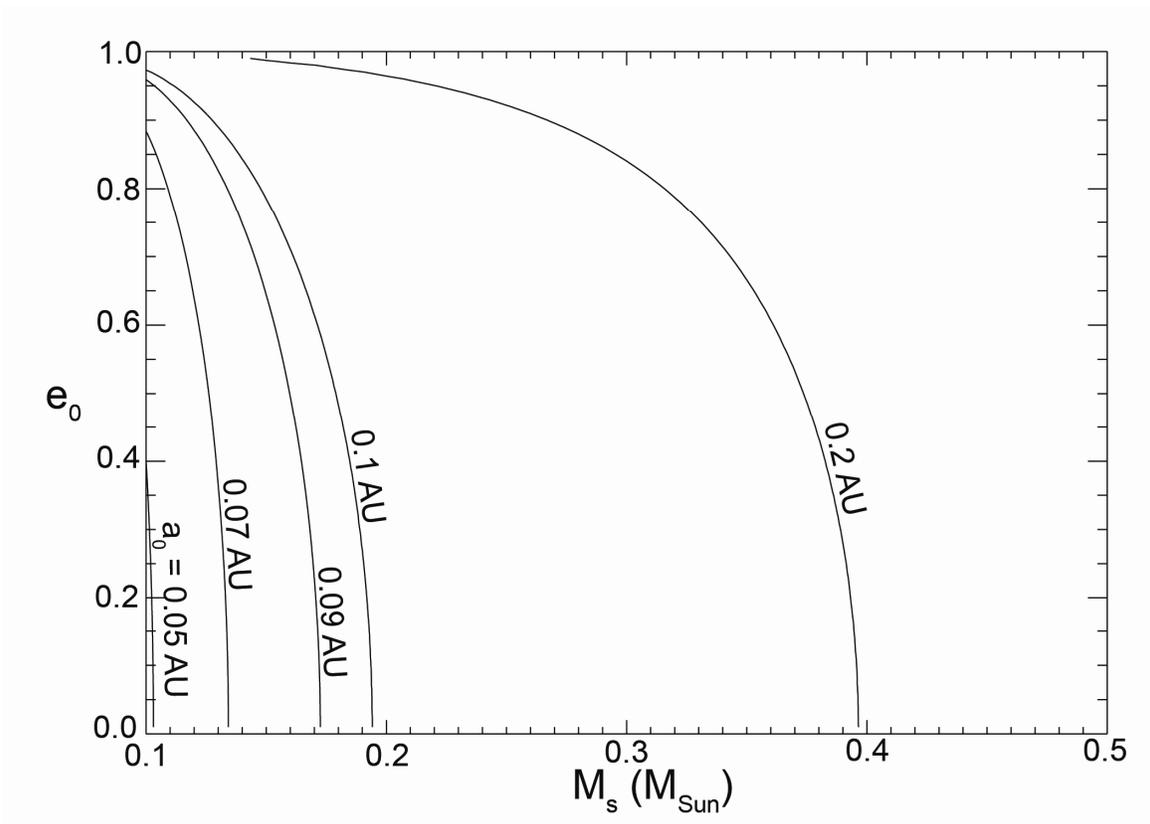

Fig. 1

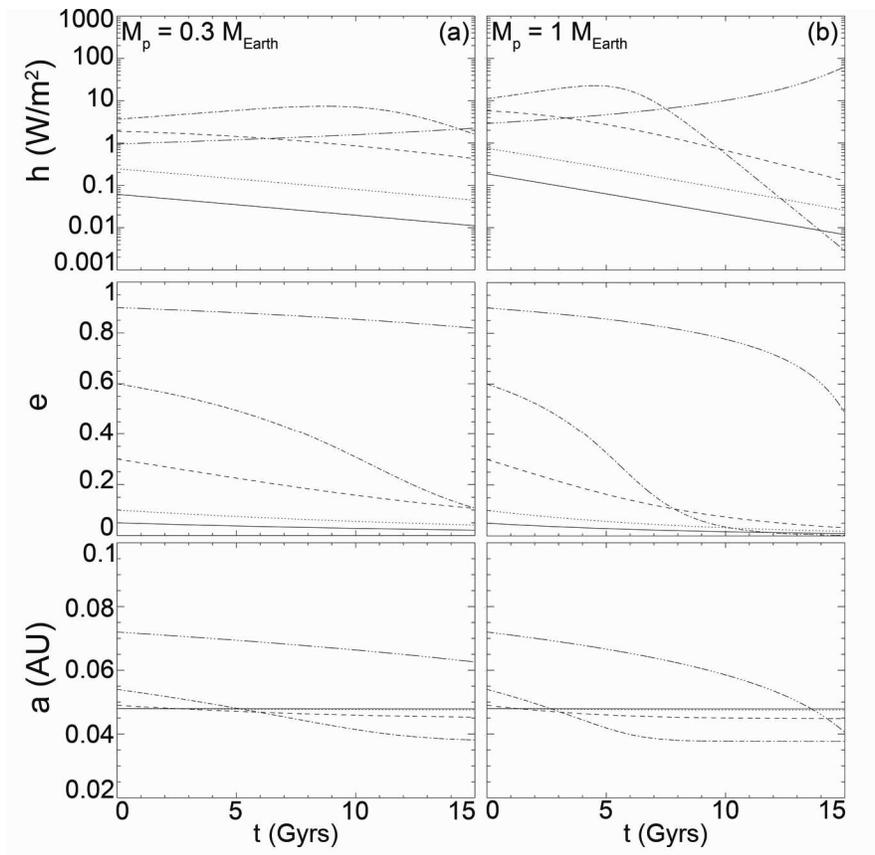

Fig. 2 (a) & (b)

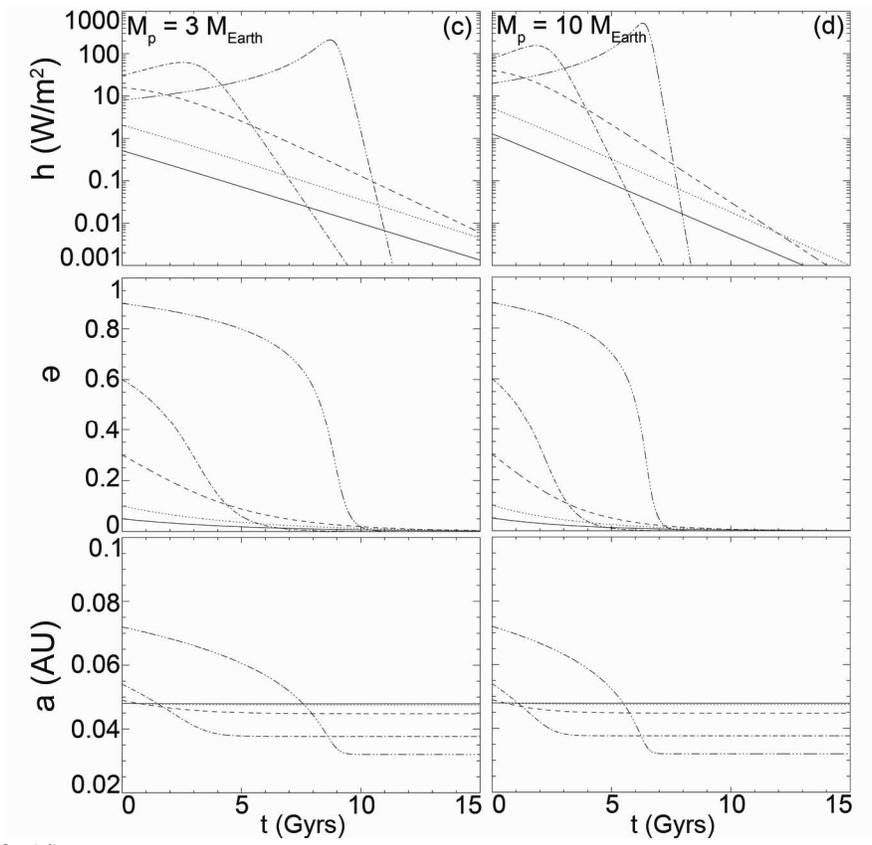

Fig. 2 (c) & (d)

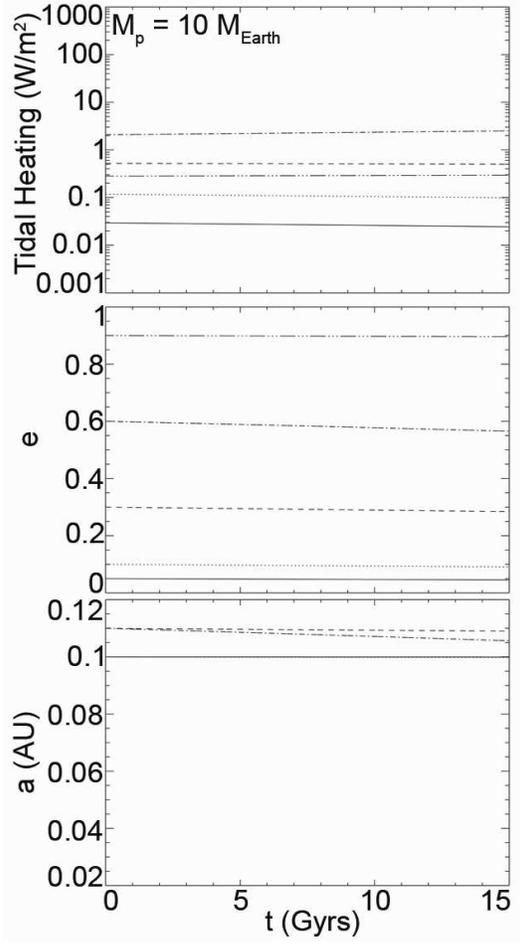

Fig. 3

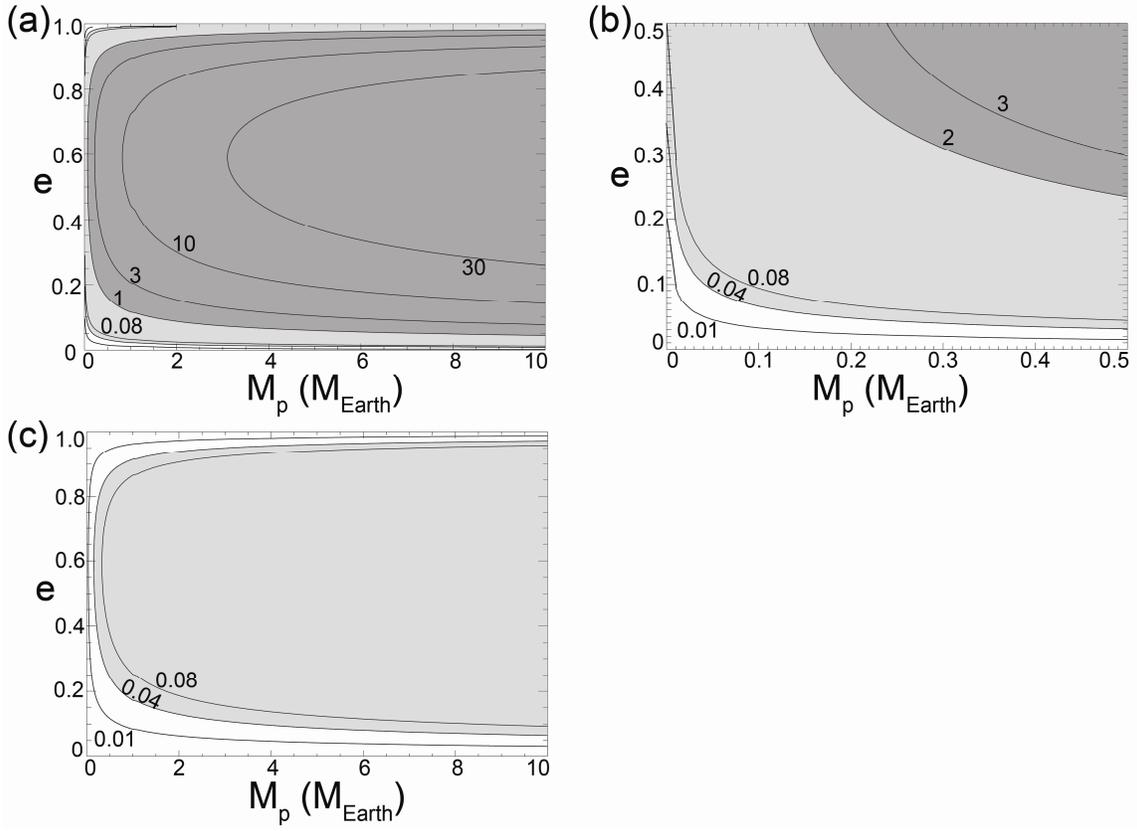

Fig. 4

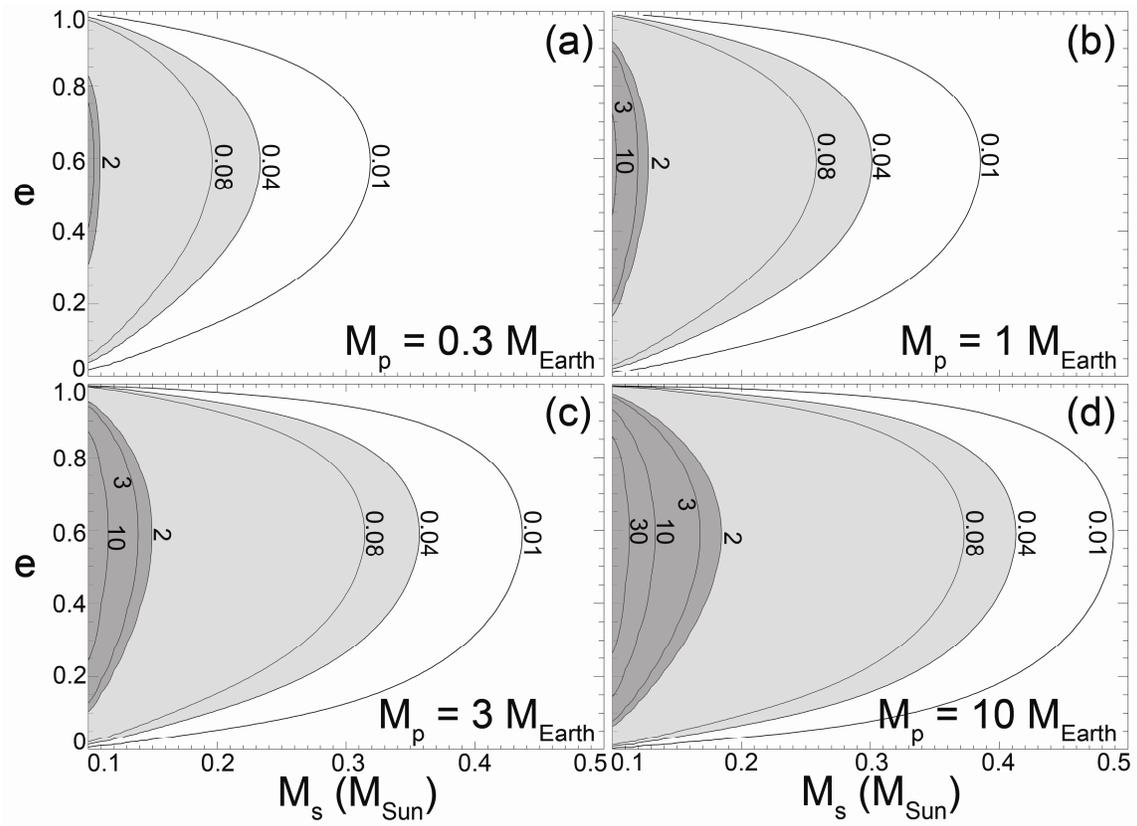
Fig. 5

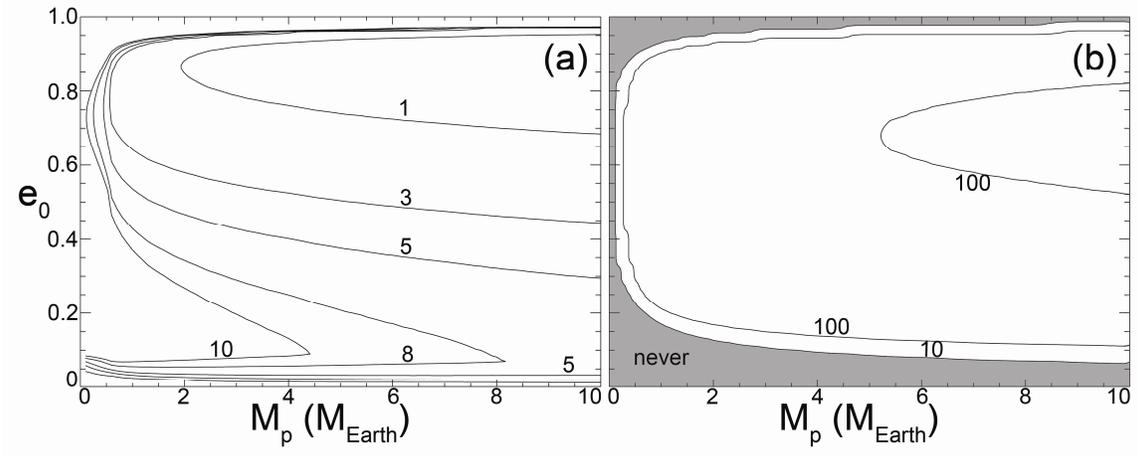

Fig. 6

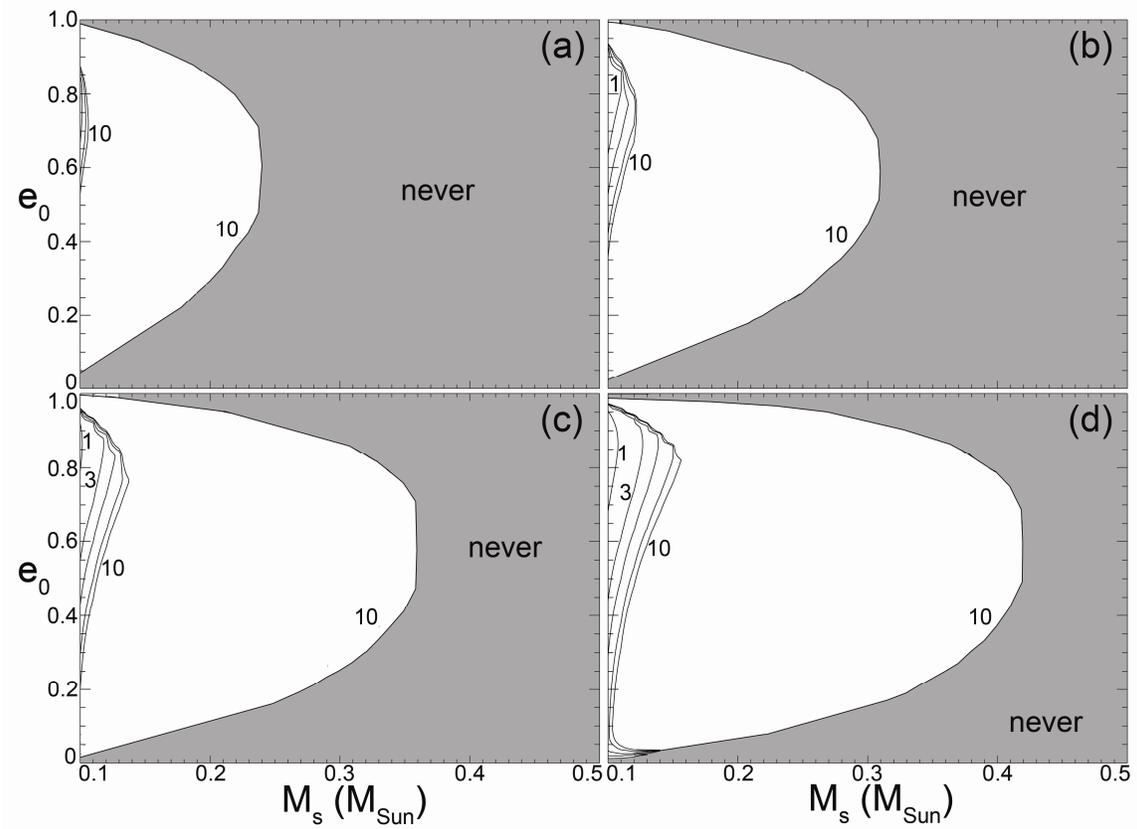

Fig. 7